\documentclass[journal,twoside]{IEEEtran}

\usepackage{enumitem}
\usepackage{lmodern}
\usepackage{listings}
\usepackage{array}
\usepackage{booktabs}
\usepackage{arydshln}
\usepackage{hhline}
\usepackage{ragged2e}
\usepackage{dirtytalk}
\usepackage{multicol}
\usepackage{setspace}
\usepackage{lipsum}
\usepackage{color}
\usepackage{multirow}
\usepackage{float}
\usepackage{csvsimple}
\usepackage{pifont}
\usepackage{algorithm}
\usepackage[noend]{algpseudocode}
\usepackage[caption=false,font=normalsize,labelfont=sf,textfont=sf]{subfig}
\usepackage{stfloats}
\usepackage{url}
\usepackage{verbatim}
\usepackage{graphicx}
\usepackage{cite}
\usepackage{orcidlink}
\usepackage{amsmath,amssymb,amsfonts}
\newlength{\firstpagerule}
\usepackage{hyperref}
\usepackage{textcomp}

\begin{document}

\title{Blockchain Powered Edge Intelligence for U-Healthcare in Privacy Critical and Time Sensitive Environment}

\author{Anum Nawaz\,\orcidlink{0000-0002-1148-0084}, Hafiz Humza Mahmood Ramzan\,\orcidlink{0009-0006-0773-3321}, Xianjia Yu\,\orcidlink{0000-0002-9042-3730}, \\Zhuo Zou\orcidlink{0000-0002-8546-1329}, Tomi Westerlund \,\orcidlink{0000-0002-1793-2694} (Senior IEEE Member)
\thanks{ Anum Nawaz is working with Shanghai Key Laboratory of Intelligent Information Processing Lab, Fudan University, China, and Turku Intelligent Embedded and Robotic Systems (TIERS) Lab, Faculty of Technology, University of Turku, Finland (e-mail: 18110720163@fudan.edu.cn). Hafiz Humza Mahmood Ramzan is working with the Department of Engineering and Architecture, University of Parma, Italy, and Xianjia Yu is a senior researcher at TIERS Lab, University of Turku, Finland }
\thanks{Zhuo Zou (Senior Member, IEEE) is a Professor with the School of Information Science and Technology, Fudan University, Shanghai, China. (e-mail: zhuo@fudan.edu.cn)}
\thanks{Tomi Westerlund (Senior Member, IEEE) is a Professor with Robotics and Autonomous Systems at the University of Turku. He leads the TIERS Lab (tiers.utu.fi), Faculty of Technology, University of Turku, Finland. (e-mail: tovewe@utu.fi)}
}

\maketitle

\begin{abstract}
Edge Intelligence (EI) serves as a critical enabler for privacy-preserving systems by providing AI-empowered computation and distributed caching services at the edge, thereby minimizing latency and enhancing data privacy. The integration of blockchain technology further augments EI frameworks by ensuring transactional transparency, auditability, and system-wide reliability through a decentralized network model. However, the operational architecture of such systems introduces inherent vulnerabilities, particularly due to the extensive data interactions between edge gateways (EGs) and the distributed nature of information storage during service provisioning. To address these challenges, we propose an autonomous computing model along with its interaction topologies tailored for privacy-critical and time-sensitive health applications. The system supports continuous monitoring, real-time alert notifications, disease detection, and robust data processing and aggregation. It also includes a data transaction handler and mechanisms for ensuring privacy at the EGs. Moreover, a resource-efficient one-dimensional convolutional neural network (1D-CNN) is proposed for the multiclass classification of arrhythmia, enabling accurate and real-time analysis of constrained EGs. Furthermore, a secure access scheme is defined to manage both off-chain and on-chain data sharing and storage. To validate the proposed model, comprehensive security, performance, and cost analyses are conducted, demonstrating the efficiency and reliability of the fine-grained access control scheme. 
\end{abstract}


\section{Introduction}
Ubiquitous healthcare refers to the availability and accessibility of healthcare services and information wherever and whenever needed, especially in situations where time is of the essence~\cite{bhatt2023improving}. It revolves around the continuous collection, transmission, and quick and seamless analysis of health-related data for making timely decisions and delivering prompt care through digital devices. This can involve technologies like real-time monitoring devices, instant communication channels, and immediate access to relevant medical data such as vital signs, physiological parameters, and various health metrics~\cite{razdan2022internet, yoon2020flexible}. Due to third-party dependencies, ubiquitous healthcare systems are still vulnerable regarding privacy and security~\cite{huang2023security,hathaliya2020exhaustive}. These cloud service providers often formulate terms and conditions primarily serving their interests. Consequently, patients (data owners) have little discretion and are compelled to accept these conditions. According to Cisco's annual report (2018–2023)~\cite{cisco2023}, 94\% of data processing is done at cloud servers due to the full offloading of data to insecure servers and continuous tracing of health data, which opens doors to privacy vulnerabilities of data owners~\cite{liang2022privacy, alzoubi2022blockchain}.

Privacy is a vital aspect of ubiquitous healthcare systems (UHS), especially in contexts where sensitive health information is involved. Ensuring the confidentiality and protection of patients' personal and medical data is essential. New regulations, like the General Data Protection Regulation (GDPR), expand the definition of personal data~\cite{godyn2022analysis}. It means that more information must be given about how data is collected and used, but data producers must be able to keep their information private~\cite{GeneralDataProtection2021}. Moreover, as users become more aware of their personal data protection laws, they demand secure frameworks from their service providers. Along with privacy issues, health data must be immutable in nature so that one can rely on these records to make medical decisions.

To solve time-related issues, researchers introduced fog and edge computing \cite{minh2022edge}. This means processing data fast right where it's needed, instead of sending all the data to servers. This quick decision-making at the edge gateways (EGs) using ML/DL helps with immediate first aid during emergencies~\cite{sharif2023priority}. Integration of edge computing with blockchain-based distributed ledger technologies (DLTs) provides efficient, robust, and reliable systems~\cite{mamun2022blockchain}. DLTs have become crucial due to the evolution of advanced security primitives and the growing demand for personalized healthcare. The introduction of DLTs offers remedies to numerous longstanding challenges within the healthcare system~\cite{zaabar2021secure,hannan2023blockchain}. 
Integration of blockchain technologies with edge intelligence for ubiquitous healthcare provides privacy, security, transparency, and streamlining of automatic processes~\cite{deep2024novel}. By using a tamper-proof and unchangeable blockchain, the authenticity of data is ensured. 

Despite inherent encryption techniques, blockchain-based healthcare applications may still risk disclosing patient information due to the possibility of linking related data on a public blockchain~\cite{belen2023systematic}. Consequently, this realm remains nascent, demanding further exploration and development efforts.

Our proposed framework, edge-gateways based ubiquitous healthcare system (\textit{EGBUH}), explored edge computing along with blockchain as a service to provide reliable, transparent, and access-controlled UHS while ensuring the anonymity of data producers. Our contributions include:
\begin{enumerate}[label=\roman*).]
    \item An autonomous computing model and its interaction topologies for privacy-critical and time-sensitive health applications.
    \item We propose a resource-efficient 1-directional convolutional neural network (1D-CNN) for multiclass classification of arrhythmia.
    \item  Continuous monitoring system, alert notification, disease detection system, data processing and aggregation, data transaction handler, and privacy handling at EGs. 
    \item Access Scheme is defined for off-chain and on-chain data sharing and storage.
    \item We conduct a security, performance, and cost analysis to demonstrate the efficiency and reliability of our proposed fine-grained access scheme.
    \item  The post-quantum key encapsulation method and digital sharing schemes for authenticity are incorporated.
\end{enumerate}
\vspace{12pt}
\section{Related Work}
Blockchain provides a safe and decentralized healthcare ledger where patient information can be securely stored and shared among healthcare professionals, patients, and providers~\cite{hemamalini2024artificial}. In~\cite{zhu2023investigation}, authors present a blockchain system model for healthcare data storage, encryption, and trading mechanisms. It ensures user data privacy by restricting storage to local spaces and decentralized networks. The personal data protection platform is implemented via code to validate the proposed theory. Authors in~\cite{monga2022mrbschain} presents a method focusing on patient-centered and private access secured by advanced encryption by utilising blockchain. It evaluates their proposed, MRBSChain's efficiency using 13 factors, comparing it to Ethereum and Binance smart chain. It also measures the speed of their proposal against a few other DLT systems.

Blockchain-powered tensor meta-learning driven healthcare system is proposed in~\cite{9973370}, leveraging IoT for secure data sharing and model training. The system uses a tensor prototype graph network for efficient modeling of heterogeneous healthcare data, ensuring strong consistency and privacy protection. Leveraging blockchain at the edge gateways provides temper-proof analytics and information management closer to the devices generating the data, reducing latency and improving response times~\cite{10053481}. Their combination can significantly enhance the way resources are used, particularly in terms of network, computing, storage, and security~\cite{XUE2023307}. 

Integration of quantum techniques and DLTs opens new doors for data privacy and security demands. Recently authors in~\cite{selvarajan2023quantum}, proposed a quantum-based consultative transaction key generation and management technique that enables secure healthcare data sharing by generating unique key pairs using random values, multiplicative operations, and timestamps. This approach enhances patient-healthcare communication and verifies users during transmission, providing a significant contribution to healthcare cyber security. In another study~\cite{wang2022quantum}, Wang \textit{et. al} suggest a blockchain algorithm combining asymmetric quantum encryption and a stake vote consensus algorithm.  The proposed algorithm in this study uses a delegated proof of stake consensus algorithm and quantum digital signature technology to secure transactions and resist quantum computation threats.
\vspace{12pt}
\section{Proposed Framework}
To enhance the experience of UHS, we proposed and developed a framework \textit{EGBUH}, an edge intelligence  (implementation of machine learning and deep learning algorithms on edge devices) based topology using ethereum as a distributed ledger technology to boost the experience of UHS. This proposed framework emphasizes early warning systems in critical situations, continuous monitoring, and personal data storage to build patient history, enabling privacy and P2P data sharing without using a third party. This hybrid UHS based on EGs allows the extension of a private blockchain based on ethereum to resource-constrained edge devices. EGs are able to define access control rules for their data, such as data trades and sharing with third parties. 
Additionally, it focuses on data producers' ownership rights and data owners' privacy. In our previous paper, we described complete implementation of ethereum network and pseudo algorithms to depict system flow \cite{nawaz2020edge}. 
\begin{table} [t]
\centering
\caption{Computing Components}
\footnotesize
\begin{tabular}[t]{@{}p{0.13\linewidth} | p{0.80\linewidth}@{}}
\toprule
 \textbf{Comp} & \textbf{Explanation}  
\\ \midrule
Monitor & This component, located closest to the sensing tier, is responsible for acquiring and aggregating data from various sources.
\\ \midrule
Analyze  & The Analyze component processes and models the acquired data, extracting meaningful insights and patterns.
\\ \midrule
Plan & Based on the analysis performed by the previous component, the Plan component constructs a procedure or strategy for the system to follow.
\\ \midrule
Execute & The Execute component is responsible for implementing the planned procedure and executing necessary changes in the system to achieve the desired outcome.
\\ \bottomrule
\end{tabular}
\end{table}

\begin{figure} [t]
\centerline{\includegraphics[width=0.98\columnwidth]{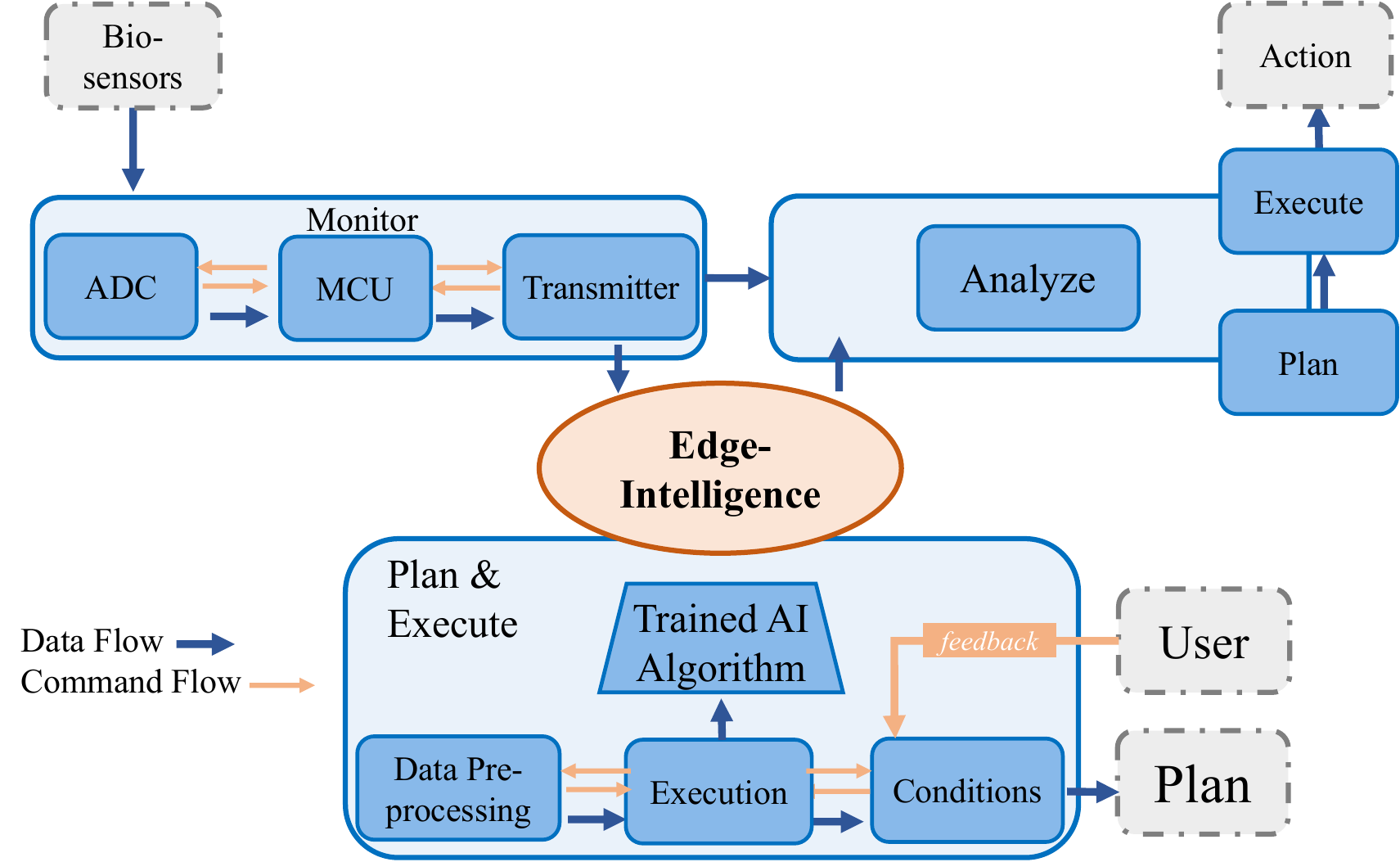}}
\caption{Components of Computing Model. a. Monitor component, b. Analyze components and c. Plan \& Execute component}
\label{fig:computingcomponents}
\end{figure}

\subsection{Proposed Computing Model}
Our proposed computing model leveraging MAPE-K (Monitor-Analyze-Plan-Execute over a shared Knowledge)~\cite{jahan2020mape} as an established framework to facilitate automated management and self-adaptive behaviour originally introduced by IBM.  This model consists of four distinct computing components, all with access to a shared knowledge base. These components are described in Table 1.
The MAPE-K model facilitates efficient automated management and adaptive behaviour within distributed systems by incorporating these four components and enabling access to shared knowledge. The hierarchical IoT architecture uses the four computing components to enable its functionality. We propose an additional component called edge intelligence to the computing model using EGs-based intelligence to implement a closed-loop technique. The role of the edge-intelligence component is to dynamically reconfigure the system's settings based on feedback received from the user's condition. A view of the enhanced MAPE-K computing components is shown in Figure~\ref{fig:computingcomponents}.

Within the framework, the perception layer encompasses the Monitor component, which operates as an intermediary bridging the sensors and other computational entities. The gateway layer comprises three distinct components: Plan, designated for localized decision-making; Execute, aimed at configuring system behaviour according to pre-decided situations; and System Management, dedicated to refining system configurations. Therefore, this entity is responsible for training an inference model, referred to as a hypothesis function, derived from user-generated data. 

\begin{figure}[h]
\centerline{\includegraphics [width=0.98\columnwidth]{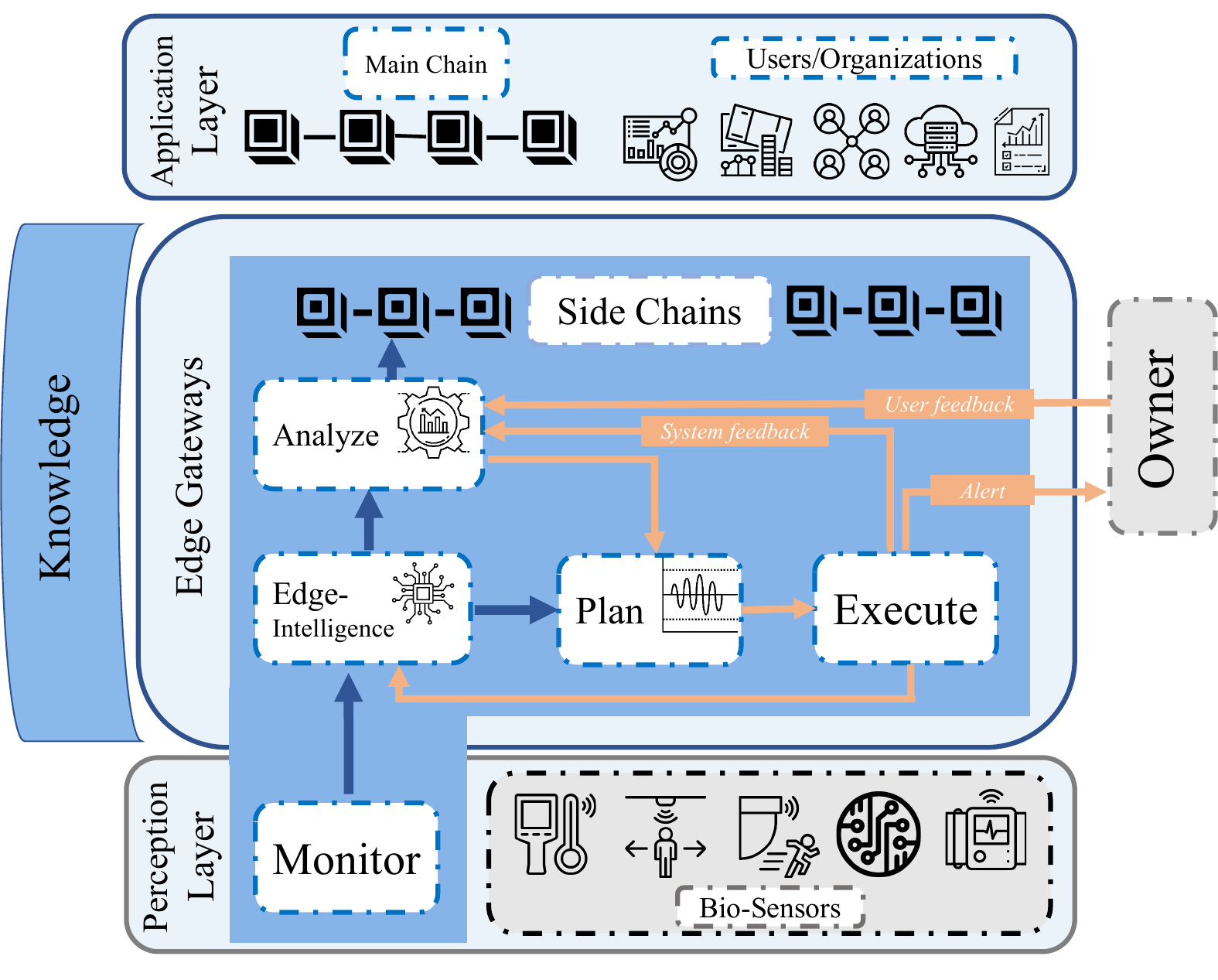}}
\caption{\textit{EGBUH}, a proposed MAPE-k based computing model}
\label{fig:computingmodel}
\end{figure}

The \textit{Plan component} implements local decisions and establishes the system's procedures. The initial step involves processing the streaming data received from the \textit{System Management} module, including feature extraction. The \textit{Execute component}, on the other hand, is in charge of actuating the system and providing feedback to other units. Upon detection of any abnormalities, users are promptly notified. These notifications are also dispatched to patients and healthcare providers as part of this process. Subsequently, the execute command proceeds to update the system management component. This allows the system's configuration to be adjusted based on the user's current state. In conclusion, the computing component provides feedback to the analysis component, such as a report of the local decisions. This feedback is used to retrain the classifier.
Distributed ledger scheme Ethereum is utilised at EGs level along with its turing complete language to incorporate user instructions at the EG level. Smart contracts convey terms and conditions based on situation and threshold values. Side chains \cite{singh2020sidechain} are utilised along with a private Ethereum network for data storage to enhance data reliability. Side chains work at EGs and save only hashes of data blocks and metadata at a local level while complete blocks are stored only in distributed data storage locations in an encrypted form using public key encryption. The discrete functions and roles of these computational components embedded within the architecture are elucidated in Figure~\ref{fig:computingmodel}.

\subsection{Proposed Access Scheme}
To implement secure communication based on TLS 1.3, we incorporate the \textit{Kyber512} \cite{avanzi2017crystals} key encapsulation method and \textit{Dilithium2} \cite{lyubashevsky2020crystals} as the digital signature algorithm for a robust and secure approach to sharing private keys of encrypted data blocks. This scheme leverages the strengths of cryptographic techniques to ensure the confidentiality, integrity, and authenticity of private keys while allowing authorized nodes to access them when needed. \textit{Kyber} is known for its post-quantum security and plays a pivotal role in this access scheme by encapsulating private keys. When a node wants to access a private key, \textit{Kyber} generates a secure encapsulation of that key, which is then transmitted over the network. The encapsulation process ensures that even if an eavesdropper intercepts the communication, it cannot derive any meaningful information about the private key without the appropriate decryption key, which remains securely stored on the target node.

\textit{Dilithium2}, on the other hand, is employed as the digital signature algorithm to authenticate the access request. When a node seeks access to a private key, it sends a request accompanied by a digital signature generated using \textit{Dilithium2}. This signature serves as proof of the authenticity and authorization of the request, making it highly resistant to forgery or tampering.

The recipient node, upon receiving the request and the associated \textit{Dilithium} signature, can then verify the signature's validity using the corresponding public key. If the signature is valid, the requestor is authorized to access the private key encapsulated by \textit{Kyber}. Only then is the encapsulated private key decrypted and made available by the requesting node. This combined use of \textit{Kyber} and \textit{Dilithium} ensures a multi-layered security approach. \textit{Kyber} safeguards the confidentiality of the private key during transmission, while \textit{Dilithium} guarantees the authenticity and authorization of the access request. The scheme provides a robust solution for securely sharing private keys within a networked system, even in the face of potential threats, thereby bolstering the overall security and integrity of the network.
\vspace{12pt}
\section{Proposed resource-optimized 1D-CNN}
We proposed resource-optimized one directional convolutional neural network (1D-CNN) architecture for the ECG Arrhythmia classification at the edge device level.  This architecture comprises two main components: the extraction phase and the classification phase. The extraction phase encompasses batch normalization, convolution, activation, and max-pooling layers, while the classification phase is characterized by flattened, fully-connected, and softmax layers.

The 1D-CNN architecture accepts an input matrix of dimensions $M*N$, where $M$ represents the length of the time window under consideration, and $N$ denotes the number of ECG channels. The initial step involves applying a batch normalization layer, which aims to standardize the input data by minimizing internal covariate shifts. Each 1D convolutional layer employs a kernel of variable dimensions $Q*N$, where $Q$ signifies the temporal window that the filter covers. These kernels move exclusively along the elements of a single dimension of the input pattern. In the 1-D convolution layer, each neuron is connected to a local window from the preceding layer, referred to as the receptive field, which shifts along the time axis and shares synaptic weights. The mathematical representation of a 1D convolutional layer is as follows:

\begin{equation}
y_r=f\left(\sum_{q=1}^Q \sum_{n=1}^N w_{q n} x_{r+q, r+n}+b\right)
\end{equation}

where \(y_r\) is the output of unit $r$ of the filter feature map of size $R$ 
(R equals to M in the case where stride=1), $x$ is the two-dimensional input portion overlapping with the filter, $w$ is the connection weight of the convolutional filter, $b$ is the bias term, and $f$ is the activation function of the filter, which in this case is $reLu$. 

This model facilitates the reduction of the number of weights and aids in the generalization process. The neurons oriented vertically represent the evolution of the input data over time, which is dependent on the receptive field and delay values. The number of neurons along the horizontal axis can be manually defined, enabling the transformation of input features into a higher-order sequence. For each neuron, the rectified linear unit function (reLu) is applied to return the weighted sum of the input data if it is positive and zero otherwise. 

To calculate the dimension of the filter feature map after the convolution operation $(R)$, the following formula is utilised:

\begin{equation}
R=\left[\frac{M-(K-1)+2}{S}\right]
\end{equation}

$S$ is the stride (the number of positions skipped by each shift of the filter during convolution).

\begin{figure}[t]
    \centering
    \includegraphics[width=\columnwidth]{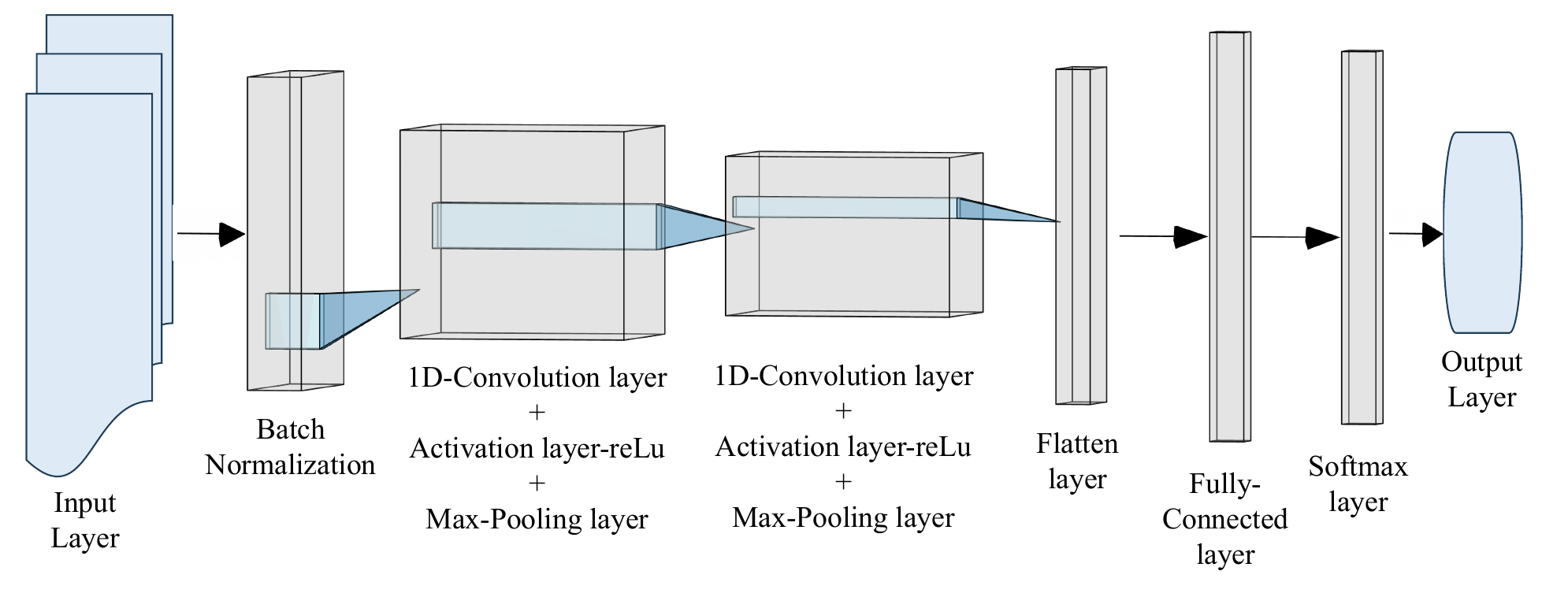}
    \caption{Optimized resource efficient 1-D Convolutional neural network}
    \label{fig:1DCNN}
\end{figure}

The overall features of the 1DCNN architecture are depicted in Figure~\ref{fig:1DCNN}. Furthermore, batch normalization ($BN$) is applied, which involves normalizing the input to the next layer, typically leading to a significantly increased learning speed and notable regularization effects that enhance the network's generalization. $BN$ operates differently during training and testing. During training, $BN$ normalizes and zero-centers the input based on the entire batch, allowing the model to learn the optimal scaling of the input. 

To normalize and zero center, the input $BN$ estimates the parameter-dependent mean $mu$ and variance $sigma^2$ computed over the batch. The zero-centered normalized value $hat{X}(i)$ for each instance is computed as $xi = 10^{-5}$ to avoid zero divisions. $BN$ adds a further step during training, using trained parameters, to further scale and offset the values as needed. 

During testing, the mean \(\mu\) and variance \(\sigma^2\) parameters cannot be computed based on the batch, so the algorithm uses the values computed with a moving average during training. 

\begin{equation}
\begin{gathered}
\boldsymbol{\mu}=\frac{1}{b} \sum_{i=1}^b \mathbf{X}^{(i)} \\
\boldsymbol{\sigma}^2=\frac{1}{b} \sum_{i=1}^b\left(\mathbf{X}^{(i)}-\boldsymbol{\mu}\right)^2
\end{gathered}
\end{equation}

In the context of $BN$, the number of instances in the batch is denoted by \(b\), and each instance is represented by \(X(i)\). Following the computation of the zero-centered normalized value \(\hat{X}(i)\) for each instance, $BN$ introduces an additional step during training. This step utilizes trained parameters to further scale and offset the values as required. 

\begin{equation}
\hat{\boldsymbol{X}}^{(i)}=\frac{\boldsymbol{X}^{(i)}-\boldsymbol{\mu}}{\sqrt{\boldsymbol{\sigma}^2+\xi}} .
\end{equation}

The element-wise multiplication, denoted by $\otimes$, involves multiplying each input value by the corresponding scaling parameter, denoted by $\gamma$. The offset parameters, denoted by $\beta$, are also learned during training. This process allows for the optimal scaling and shifting of the normalized values, enhancing the network's performance by adjusting the distribution of inputs to fall within a specific range, thereby facilitating more effective training. 

\begin{equation}
\mathbf{z}^i=\boldsymbol{\gamma} \otimes \hat{\boldsymbol{X}}^{(i)}+\boldsymbol{\beta}
\end{equation}

The second convolutional layer is with the same parameters as the first convolutional layer. Pooling is crucial for CNNs to reduce the input size, and decrease the required computation, and the number of network parameters. 

Furthermore, this size reduction tends to make the representation space invariant concerning small translations of the input, allowing the network to recognize specific patterns at different locations within the feature map. A one-dimensional max-pooling layer is applied to preserve, for each activation map, the neuron with the higher value. The classification part is analogous to a multi-layer perception.

\begin{equation}
y_j^{(l)}=f\left(\sum_{i=1}^I w_{j i}^{(l)} \cdot x_i^{(l-1)}+b_j^{(l)}\right)
\end{equation}

This is followed by a flattened layer that reshapes the matrix input into a vector to support the processing of the subsequent non-spatial layers. The flattening layer consists of converting the data of the extraction part into a 1D-vector format. One hidden layer with the dropout function is implemented, and the neurons of the output layer correspond to the classes of heartbeats disease. Each unit activation $y_j(l)$ is computed as follows:

\begin{equation}
f(x)=\left\{\begin{array}{cc}
x, & \text { if } x>0 \\
0.01 \cdot x, & \text { otherwise }
\end{array} .\right.
\end{equation}

\begin{equation}
\hat{y}_i=\operatorname{argmax}\left(\frac{e^{y_i}}{\sum_{i=1}^5 e^{y_i}}\right) .
\end{equation}

In the simplest case, each unit is retained with a fixed probability $p$ independent of the other units. The output layer nodes of the proposed model represent 5 different heartbeat groups as specified by the Association for the Advancement of Medical Instrumentation (AAMI) standard.
\vspace{12pt}
\section{Experimental Testbeds}
In this section, we implement our proposed framework \textit{EGBUH} as a continuous monitoring system of individuals along with edge-level intelligence to secure processed information. 

\begin{figure}[h]
\centering
\includegraphics[width=\columnwidth]{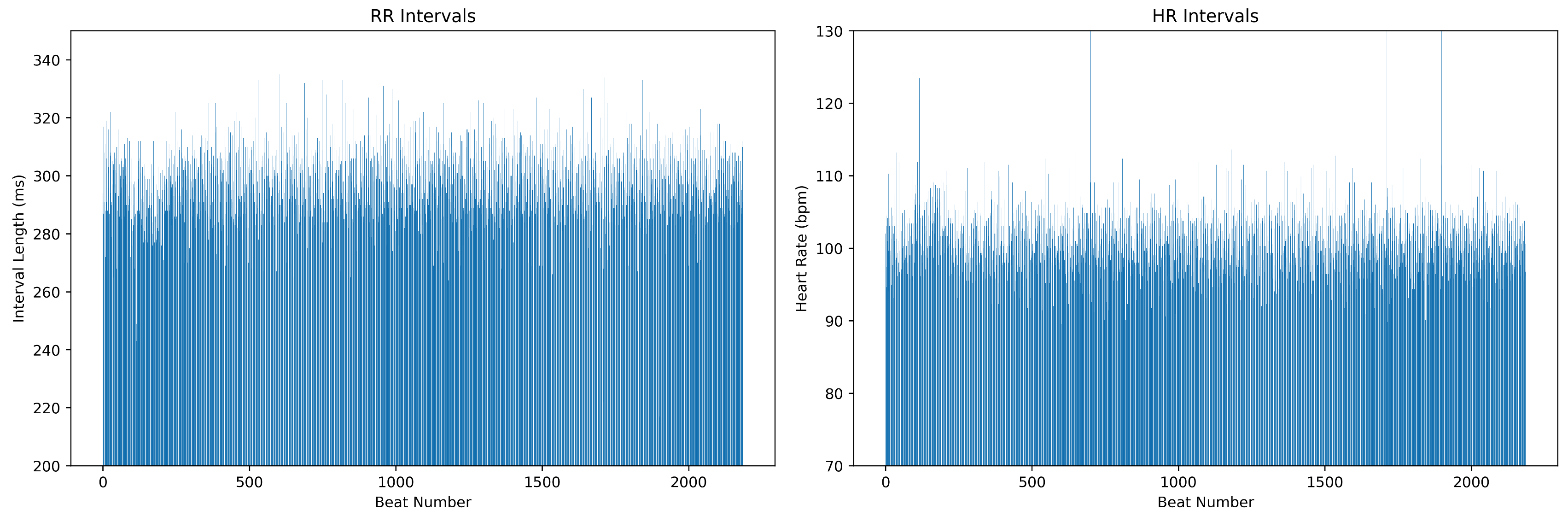}
\caption{RR and HR values from random sample of 25 minutes ECG recording with 500 Hz sampling rate and 12 bits resolution}
\label{fig:HRValuesvariations}
\end{figure}

To assess the effectiveness of \textit{EGBUH} across diverse dimensions, two machine learning algorithms are employed. Initially, abnormality detection through simple binary classification on the signals are carried out using a linear Support Vector Machine (SVM) approach. In a separate scenario, a 1D-CNN served as a deep learning algorithm to identify various arrhythmias, constituting a multi-class classification task. We compared two distinct testbeds against the baseline ECG system for reference.

\subsection{First Testbed}
In our first testbed, a 2-channel ECG system is employed, with the perception layer anchored by the AD8232, a specialized signal conditioning block tailored for electrocardiogram (ECG) and heart rate monitoring. The EGs consist of Arduino UNO, STM32F427, and Raspberry Pi 3. The sensor node AD8232 captures data and transmits it to the edge gateway node. An ECG feature extraction service is implemented within the EGs to derive crucial parameters like heart rate, P wave, and T wave from the ECG signal, progressing through several stages: movement artefact removal, wavelet transformation, threshold estimation, and P wave and T wave detection.

The movement artefact removal phase employs band-pass and moving average filters to counter environmental noise (e.g., 50 Hz power-line noise). The filtered data then enters the wavelet transformation, where the Daubechies-4 wavelet is chosen due to its efficiency in extracting P-wave and T-wave components without excessive computational delay. Thresholds for identifying R, P, and T waves are determined based on the wavelet transformation results, with R wave thresholds generally higher in millivolts compared to P and T waves. For instance, 1 $mV$ might be set as the R peak threshold in the lead I, while thresholds of 0.08 $mV$ and 0.1 $mV$ are used for P and T waves in the lead II, respectively. These values can vary depending on the specific ECG leads utilized. The heart rate is computed using the R-R interval information derived from these thresholds, applying the formula:\\

Heart rate = $60/R-R$ interval.\\

The proposed ECG feature extraction, driven by wavelet transformation, efficiently employs network bandwidth. At each discrete wavelet transform level, the data sample count is halved. Instead of storing the raw ECG data, emphasis is placed on preserving data following the wavelet transformation and the associated coefficient values. This approach significantly conserves network bandwidth between 40\% and 80\%, contingent on the wavelet transformation types and levels, while potentially introducing a minor uptick in system latency. The selection of wavelet transform types and levels should be carefully calibrated based on specific application needs to minimize potential errors in the inverse transformation process.

In essence, deploying the ECG feature extraction service within EGs, employing the outlined template and wavelet transformation methods, facilitates accurate real-time monitoring of vital parameters while resourcefully optimizing network assets.

\begin{figure}[h]
\centering
\includegraphics[width=\columnwidth]{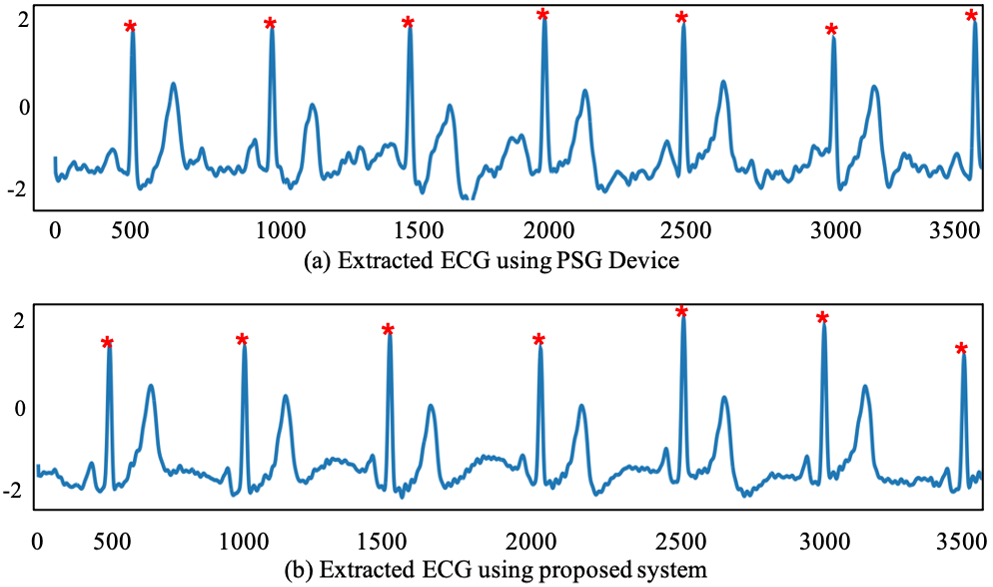}
\caption{ECG extracted cycles (a) Extracted ECG using PSG Device (b) Extracted ECG using proposed system}
\label{fig:ecgpcgproposed}
\end{figure}
\medskip

\subsection{Second Testbed} 
The second setup encompasses an ECG setup of three electrodes, the positive electrode, negative electrode, and Driven Right Leg (DRL) to detect electro potential changes due to cardiac electrical activity. 
The ADS1292r serves as the electric signal converter for capturing the ECG signal in this system. The hardware is equipped with two 32-bit microcontrollers (MCU), namely the STM32F401CCU6 and STM32F103C8T6 from ST Electronics. Additionally, it incorporates digital-to-analogue converters (DAC) MCP4921, general-purpose operational amplifiers (OPAMP) LF353, and a Wye resistor network. The STM32F401CCU6, belonging to the ARM Cortex-M4 cores family, operates at 84 MHz, offering a balance of cost-effectiveness and high performance. It features standard communication peripherals such as SPI, I2C, USB, USART, and CAN. The MCU includes a single-precision floating-point unit for rapid calculations, digital signal processing instructions, and two analog-to-digital converters. The DACs are configured to operate with an external voltage reference, receiving clock signals up to 20 MHz from the MCU. Stable DC supply powers the DACs, providing the necessary external voltage reference of +5 V.

To minimize noise impact on signal integrity, a bypass capacitor is introduced. ECG waveforms with amplitudes tenfold larger than real ECG amplitudes, including offset levels simulating baselines, are generated. The converter outputs undergo a non-inverting amplifier stage to ensure physiologically consistent amplitudes. This amplifier, comprising general-purpose operational amplifiers and resistors (R1, R2, Rg, and Rf), produces differential output signals with low offset voltage and minimal noise. The design generates ECG waveforms within the 0.5 mV to 4 mV range, exhibiting low noise and limited offset effects. Data transmission from the module occurs via the serial peripheral interface (SPI) and is sent to EGs. Three electrodes are strategically positioned: the positive electrode on the left arm, the negative electrode on the right arm, and the DRL on the right arm, with a horizontal separation of about 5 cm. The RLD electrode enhances the common mode rejection ratio (CMRR) by transmitting the common mode signal of the two sensing electrodes back to the user's body. The analogue front end integrates the ADS1292, a programmable gain amplifier (PGA), an analogue-digital converter (ADC), and an RLD circuit. Data collection duration varied based on participant age groups, dividing data into chunks of 10 minutes for young individuals and 5 minutes for elderly participants. Collection was conducted in shorter segments of 2, 5, and 10 minutes each.

Post-acquisition, data is transmitted to data transmission modules via SPI. The MCU and Wi-Fi module then transmit signals via Wi-Fi to the user interface. A Xiaomi power bank supplies power to the hardware system, utilizing the ESP32's built-in WiFi module for communication. During measurements, some subjects were supine and instructed to relax muscles, minimizing muscular artefacts and evaluating pure ECG quality. High-pass and low-pass filters with cutoff frequencies of 0.1 Hz and 200 Hz were implemented in the measurement electronics. After digitization at an 800 Hz sampling frequency, notch filters at multiples of 50 Hz were used to eliminate power line interference. Expectedly, movement introduces diverse movement artefacts in the ECG signal, particularly muscular artefacts and baseline shifts. However, these can be mitigated effectively through digital signal-processing techniques.


\section{System Implementation}

The performance of a proposed system in terms of resource utilization and time is investigated through raspberry pis and STM32 M4-cortex microcontrollers. The raspberry pi devices run raspbian, based on linux kernel version 4.14.52-v7+, with 1GB of RAM and 4-core ARM processors (BCM2837 @ 1.4GHz). The STM32 family relies on a linux system. Communication between different edge-gateways and their perception layer is facilitated through a local Wi-Fi network. Raspberry Pi 3 Model B+ and STM32 M4-cortex microcontrollers are utilized at edge-gateway layer as manager nodes. Lightweight nodes are implemented using ardiuno and less-computational powered bio-sensors.
For simulating the application layer, desktop computers with an intel core i7 processor and 16GB of RAM are employed. The proposed approach is trained and investigated using the MIT-BIH dataset and real-time ECG signals collected from 18 individuals aged 18- 52 years. 

Ethereum, a private network is leveraged as a service layer. To implement various components of the system, we utilize the Go programming language (golang), solidity for smart contracts, and a suite of web technologies (Node.js\textregistered, HTML5, CSS3, jQuery) for the front-end application. Smart contract deployment is accomplished using the Remix IDE. Metamask, a browser extension facilitating parallel transaction flows during experiments, generates data requests and transactions to and from third-party cloud services. Proof-of-stake consensus is used for block confirmation and gossip protocol to ensure fast, and attack-resilient message propagation for transaction handling, keeping nodes synchronized and avoiding forks.

\begin{figure}[h]
\centerline{\includegraphics[width=\columnwidth]{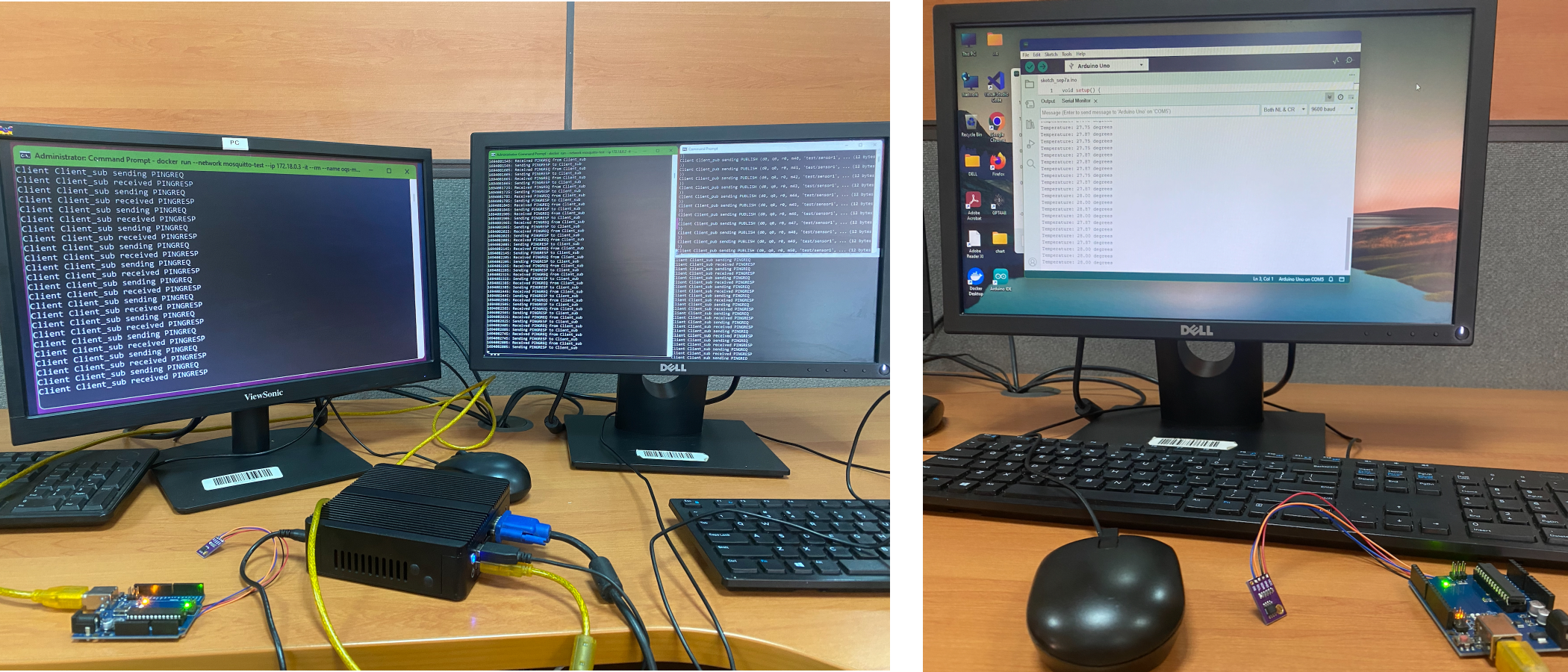}}
\caption{System Setup using Mosquitto MQTT broker}
\label{fig:kybersetup}
\end{figure}

Figure ~\ref{fig:kybersetup} shows the system setup. To read data from sensors and send it to the Mosquitto MQTT broker over a TLS 1.3-secured connection, we configure the Mosquitto MQTT broker, which involves setting up server certificates and configuring the mosquitto.conf file to use these certificates, and specifying the TLS version and port number. To run parallel ports, we use one linux-based system and assign static IP addresses on different ports by using multiple containers in docker, sensors, and ardiuno board for real-time implementation. To create an MQTT client on arduino board, we use the Arduino MqttClient library, to publish/subscribe to MQTT topics. 

\subsection{Early warning system}
The real-time push notification service serves to promptly notify designated individuals (e.g., guardians) of detected abnormalities, ensuring swift responses such as immediate first-aid interventions. This service triggers notifications upon detecting abnormal heart rates or ECG signals (e.g., prolonged P waves or elevated T wave amplitude). Additionally, notifications are dispatched if the internal temperature of a smart gateway surpasses a predefined threshold or if the gateway ceases to receive incoming data from sensor nodes over a specified time span. The content and priority level of push messages vary based on specific events. For instance, a heart rate exceeding 80 bpm triggers a priority level 1 message, while a heart rate above 120 bpm prompts a priority level 3 message. Depending on the application, push notifications can be executed and activated at the gateway level.

\begin{figure}[t]
\centering
\includegraphics[width=\columnwidth]{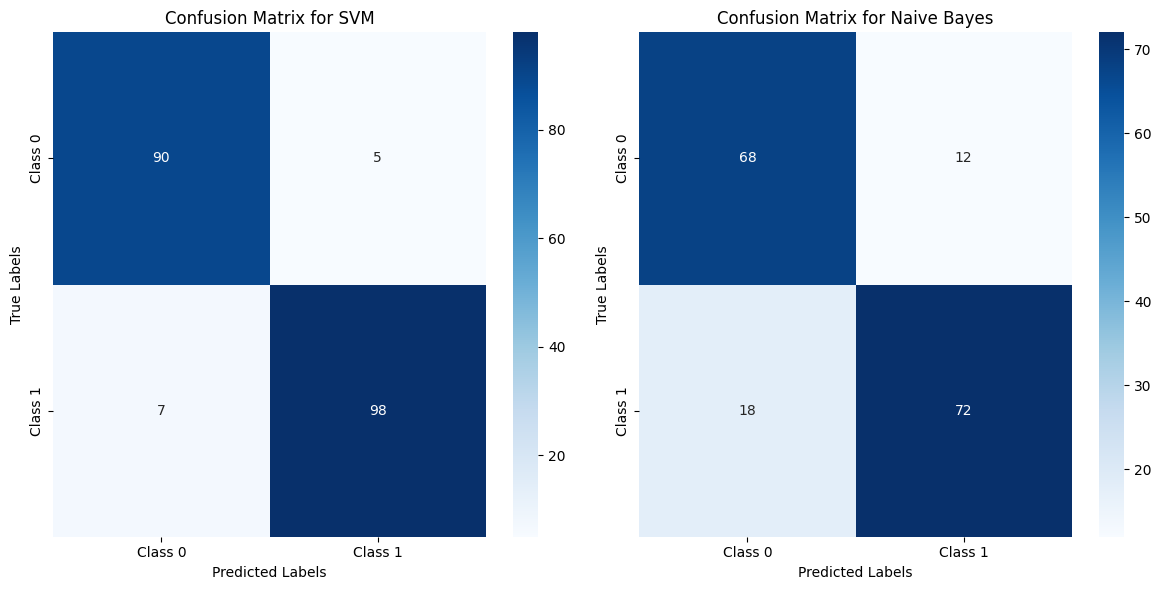}
\caption{Confusion Matrix for Abnormality Detection using ECG signal processing (a) Abnormality Detection Accuracy using SVM  (b) Abnormality Detection Accuracy using Naive Bayes}
\label{fig:SVMNaiveBayes}
\end{figure}

In the proposed systems, the push notification service harnesses binary classification, distinguishing normal and abnormal beats in the initial stage to prompt first aid actions. To validate the system's effectiveness, the model was trained using Physiobank databases \cite{physhik}, and realtime testing using collected data samples utilising python libraries such as Scikit\-learn \cite{scikitle47} and Biosppy \cite{biopsy1}. Real\-time decision-making regarding a user's health condition is enabled through the utilization of the linear Support Vector Machine (SVM) method and Naive Bayes due to their less complexity and fast response rate. Figure ~\ref{fig:SVMNaiveBayes} depicted that SVM shows better results in terms of accuracy as compared to Naive Bayes. This proposed method classifies incoming signals as either normal or abnormal. 

We collected 2 hours of ECG signals from healthy individuals and cardiovascular issues. The perception layer divides the chunks of 10 seconds of signals and sends them to EGs where data pre-processing is handled and labels the signal as normal or abnormal. These features encompassed QRS complex duration, T wave duration, RR 
interval, PR interval, and ST segment (refer to figure ~\ref{fig:ecgpcgproposed}). During runtime, incoming test data were locally classified, with the decision vector sent to the Execute component for actuation. Test data included ECG signals with random arrhythmia points added to normal ECG data to simulate emergency scenarios. These scenarios were tested on data from four new users.

\subsection{Arrhythmias Detection }
Beyond employing a linear machine learning approach, the capability of integrating a non-linear algorithm into this architecture was explored. Incorporating such algorithms opens avenues for multivariate and intricate applications within IoT-based health monitoring systems \cite{alamsyah2020internet}. A Convolutional Neural Network (CNN), a deep learning algorithm, was harnessed for real-time multi-class classification. To this end, we adopted our proposed 1-DCNN ~\ref{fig:1DCNN} and Long short-term memory (LSTM) model for multi-class ECG signal classification using the TensorFlow library in python, the algorithm underwent training in the \textit{Analyze} component ~\ref{fig:computingcomponents}). The trained hypothesis function was subsequently transmitted to the EGs, enabling the classification of incoming ECG signals into five classes: normal (N), supraventricular ectopic beat (SVEB), ventricular ectopic beat (VEB), fusion beat (F), and unknown beat (Q). Table ~\ref{tab:Samplenumbers} depicted the number of samples against each heartbeat class.

EGs receive raw data from the perception layer, pre-processed and normalized it. After normalization, apply binary classification using SVM. If the classification shows irregularities, it will generate an emergency alert to the patient and concerned bodies. After creating an alert, multi-classify using 1-DCNN. Otherwise, create a hash to save the information into a new data block. There are two parts to this data block, a header and a body. The body part of the block contains processed information and a header part that documents the general characteristics of the information. Among these factors are the hash of the previous block, the time stamp, and the type of data, which can be further processed for the purpose of intelligent systems by combining heterogeneous data at a higher level. A symmetric cryptographic scheme is used to protect the hash of the data block. A blockchain cloud saves the encrypted data blocks, and the EG holds only the key. The client (doctor, specified person) will use this private key to decrypt the desired data. In addition, it will also define the access control parameters of the data.

\begin{table}[t]
    \centering
    \caption{Number of samples against each class of heartbeat}
    \label{tab:Samplenumbers}
    \small
    \begin{tabular} {@{}p{0.36\linewidth}p{0.15\linewidth}p{0.17\linewidth}p{0.17\linewidth}@{}}
        \toprule
       \textbf{Class} & \textbf{Serial number}  & \textbf{Training Samples} & \textbf{Testing Samples}  \\
        \midrule
        Normal &  0 &  72471 & 18118\\
        Fusion of paced and normal & 1 &  2223 & 1608\\
        Premature ventricular contraction & 2 &  5788 &  1448\\
        Artial Premature & 3 &  641 & 556\\
        Fusion of ventri and normal & 4 &  6431  & 162\\
        \bottomrule
    \end{tabular}
\end{table}

\subsection{Ethereum Procedural Steps}

The experimental setup is divided into six processes, as described below. Each process is accompanied by a series of steps. $SC$ is used to refer to smart contracts.

\begin{enumerate}
    \item \textbf{System initialization:}
The process of setting up a private ethereum network involves several steps. Encryption parameters and genesis file is generated, and the network is initialized. During this phase, clients and their devices are registered to the network, a necessary step for re-authentication. The device registration phase is handled by EGs and the device registration request as a new transaction proposal. 
The generation of encryption parameters, creation of the genesis file, and initialization of the ethereum blockchain are performed through smart contracts.

\item \textbf{Generation of encryption keys:} Each connected device generates its key pair of secret and public keys $(s_k, p_k)$. The secret key is randomly generated, and then the private key and child secret keys are derived from it. Each child's secret key is used to encrypt one data batch. For example, each child's secret key is used to encrypt a specific batch of data. The complete encryption process, along with key encapsulation and digital signature algorithm, is described in the section proposed access scheme.

\item \textbf{Data processing, encryption, and storage:}  Edge gateways divide the acquired data segments into data batches after every time $T$ and implement embedded edge AI algorithms. If edge gateways do not have enough resources to process huge block of data, this specific data block is saved as raw data. After encryption, devices call the smart contract function. Each batch is sent to storage after encryption and broadcasting the hash and metadata.

\item \textbf{Transaction Initialization:} On each data request, a smart contract function is used to query available records for a data batch. The buyer can initiate trade requests for its selected data batches after pre-defined deposit agreement. To perform this action, the smart contract function is called through which the secured channel builds and sends a secret key after the key encapsulation method and digital signature algorithm. It consists of its address or identifier and the batch identifier. 
    
\item \textbf{Transaction Confirmation:}  This process runs periodically requesting any \textit{DataBatch} requests that are present and depositing the requested data. EGs encode the key using \textit{kyber512} key encapsulation method (KEM) and confirm the \textit{Deal} if any deposit meets the pre-defined conditions for the type of data in the requested batch. A combination of \textit{kyber512} and \textit{dilithium2} is used and the process is automated using turing complete capability of smart contracts. 

\item \textbf{Finalization of data transfer:} When  a receiver receives an encrypted data batch key, it uses the decryption algorithm to obtain the batch key. Then, it queries the storage provider with the batch address and decrypts it to obtain the information. The data transfer, or purchase, is done if a buyer is satisfied. If the receiver is unsatisfied with the received data batch, it will ask for a refund. The edge gateways will resolve a dispute by cross-checking the batch details. After getting the results, gateways will respond accordingly.
\end{enumerate}

\vspace{12pt}
\section{Performance Analysis}

This section presents and discusses the results of the experiments as well as the performance of the proposed model. To evaluate the possibility of real-time operation, we examined the execution times of feature extraction, arrhythmia detection, and blockchain requests. To check the overall consumption of hardware resources and efficiency of EGs, two light-weight algorithms, LSTM and 1-D CNN, were trained and tested on MIT-BIH databases\cite{physhik} to train and test the resource consumption and accuracy of EGs. 

EGs receive raw data from the sensor nodes. Raw ECG signal data is pre-processed and normalized. The extraction includes batch normalization, convolution, activation, and max-pooling layers. The flattening, fully connected, and softMax layers constitute the classification part. Input is standardized using a batch normalization layer to reduce the internal co-variate shifts. Each neuron is connected to the local window from the previous layer, known as the receptive field, which shifts according to timestamps and shares synaptic weights. By using this approach, we reduce the number of weights, which facilitates the generalization process. A Rectified linear unit function (ReLu) is applied to return the weighted sum of the input data. After this, a one-dimensional maxpooling layer is applied to preserve the neurons of each activation layer. However, the classification part is the same as with multi-layer perceptron. Table ~\ref{tab:CNNconfiguration} presents the CNN network configuration used to train and test arrhythmia classification.

\begin{table}[h]
\begin{center}
\caption{1D-CNN Network configuration}
\label{tab:CNNconfiguration}
\small
\begin{tabular}{@{}p{0.30\linewidth}p{0.25\linewidth}p{0.33\linewidth}@{}}
\hline \textbf{Network Part} & \multicolumn{2}{|l}\textbf{{ Description } }\\
\hline \hline \multirow{5}{*}{ Extraction part} & \multicolumn{2}{|l}{ 319 Neurons } \\
\hline \multirow{5}{*}{ Input Layer } & Conv-1D & \begin{tabular}{l} 
Kernels: 64 \\
Receptive field: 2 \\
Stride: 1\\
\hline
\end{tabular} \\
 & Activation & reLu \\
 & Dropout & Probability: 0.4 \\
 & Max-pooling & Pool size: 2 \\
\hline \multirow{3}{*}{ Classification part } & FC layer & 512 neurons \\
 & Dropout: & Probability: 0.2 \\
 & Output layer: & 5 neurons \\
\hline
\end{tabular}
\end{center}
\end{table}

The baseline LSTM model ~\cite{khan2021cardiac}  is also used to train and predict the raw data at the EGs to compare its computational efficiency with our proposed 1D-CNN. Usually, LSTM is preferred for edge-based systems due to its lower resource consumption and better fit for sequential data. LSTM enables the system to forget about unnecessary information from the previous outputs, which makes it suitable for scarce computing devices. After this, new input $X(t)$ is decided, and apply the sigmoid function to decide the updation of the next value. A $tanh$ layer creates the vector of all possible values from the upcoming input. The sigmoid layer decides the part of the information that will go to the final layer. The trained hypothesis function was subsequently transmitted to the EGs, enabling the classification of incoming ECG signals into five classes: 0=Normal, 1=Fusion of paced and normal, 2=Premature ventricular contraction, 3=Artial Premature, 4=Fusion of ventri and normal.

The average performance measures of 1D-CNN are described in Table ~\ref{tab:1DCNNperformance} and an average performance matrix of LSTM is described in Table ~\ref{tab:LSTMPerformancemeasures} respectively. From the comparison results in table ~\ref{tab:1DCNNperformance} and table ~\ref{tab:LSTMPerformancemeasures}, it can be concluded that the proposed 1D-CNN performs better in comparison to the baseline LSTM model while utilising the same resources and time. A comparative analysis of our proposed 1D-CNN with similar studies demonstrates an average accuracy of 97.4\%, while utilizing significantly fewer resources than other studies.

\begin{figure}[t]
\centering
\includegraphics[width=\columnwidth]{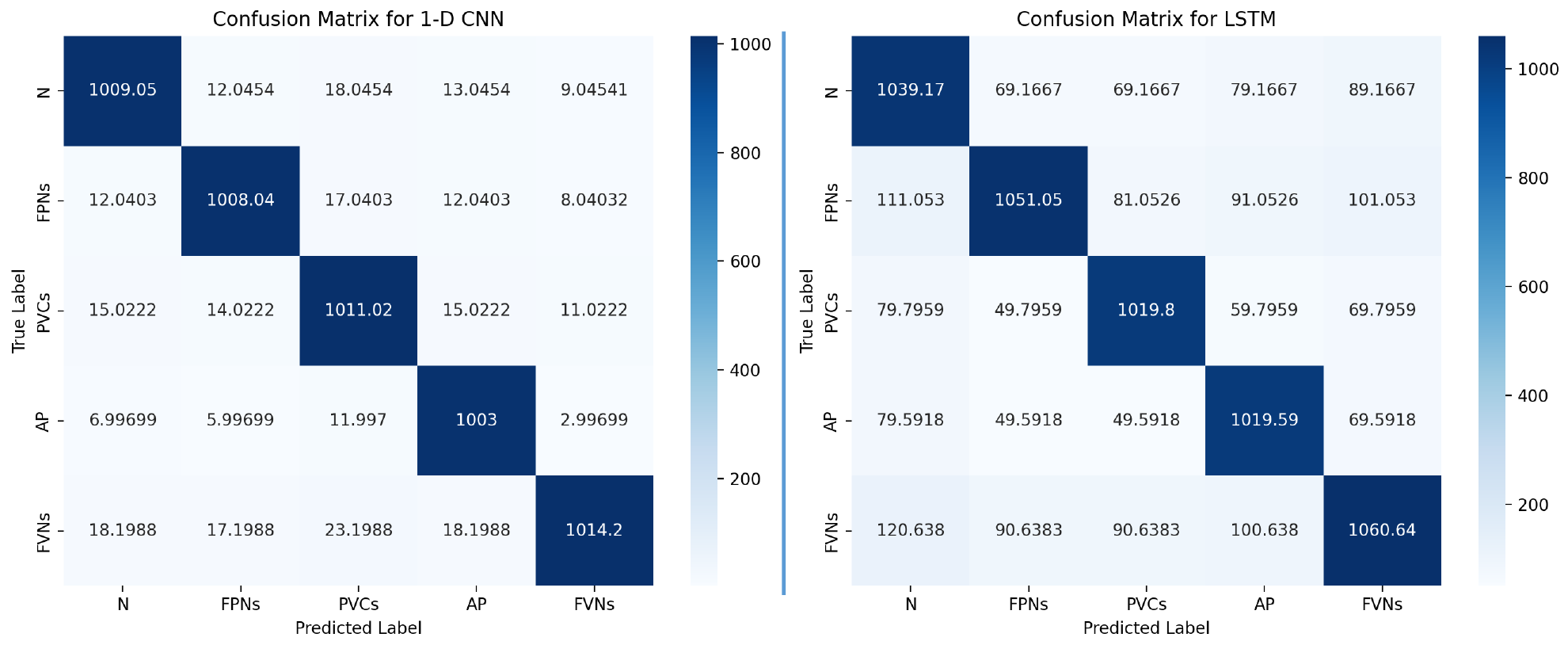}
\caption{Confusion Matrix of proposed 1D-CNN and LSTM baseline model}
\label{fig:cmfinal}
\end{figure}

Results of the confusion matrix in figure ~\ref{fig:cmfinal} demonstrate high accuracy of 1D-CNN, with most predictions correctly placed on the diagonal, indicating strong performance and minimal misclassifications using a balanced dataset. The limited off-diagonal values suggest that the model rarely confuses one class with another. In contrast, the LSTM model, while still performing well, has slightly more off-diagonal values, indicating a higher number of misclassifications. This is reflected in its slightly lower precision and recall compared to the 1-D CNN, with the confusion matrix showing a bit more spread, indicating that the LSTM is less certain in its predictions.

\begin{table}[t]
    \centering
    \caption{Average Performance measures of proposed 1D-CNN}
    \label{tab:1DCNNperformance}
    \small
    \begin{tabular}{@{}p{0.20\linewidth}p{0.15\linewidth}p{0.15\linewidth}p{0.15\linewidth}p{0.13\linewidth}@{}}
        \toprule 
     \textbf{Category} & \textbf{Accuracy} & \textbf{Precision}  & \textbf{Recall}  & \textbf{f1 score}  \\
        \midrule
        \midrule
           N        &  0.996    &   0.991    & 0.996    &  0.994  \\
           FPNs     &  0.981    &   0.992    & 0.997    &  0.991  \\
           PVCs     &  0.990    &  0.989     & 0.991    &  0.994  \\
           AP       &  0.989    &   0.997    & 0.996    &  0.990  \\
           FVNs     &  0.998    &  0.986     & 1.000    &   0.999 \\
           macro avg            &  0.9908     & 0.991    &  0.996   &   0.9938 \\
           wghtd avg         &  0.9904     &   0.9852  & 0.9911    &   0.9895  \\
        \bottomrule
    \end{tabular}
\end{table}

\begin{table}[t]
    \centering
    \caption{Average Performance measures of baseline LSTM \cite{khan2021cardiac}}
    \label{tab:LSTMPerformancemeasures}
    \small
    \begin{tabular}{@{}p{0.20\linewidth}p{0.15\linewidth}p{0.15\linewidth}p{0.15\linewidth}p{0.13\linewidth}@{}}
        \toprule 
     \textbf{Category} & \textbf{Accuracy} & \textbf{Precision}  & \textbf{Recall}  & \textbf{f1 score}  \\
        \midrule
        \midrule
    N       & 0.94 & 0.96 & 0.94 & 0.94 \\
    FPNs    & 0.96 & 0.95 & 0.97 & 0.96 \\
    PVCs    & 0.96 & 0.98 & 0.97 & 0.96 \\
    AP      & 0.97 & 0.98 & 0.96 & 0.96 \\
    FVNs    & 0.94 & 0.94 & 0.95 & 0.93 \\
    macro avg     &  0.9908   &  0.9910    &    0.9960 & 0.9923   \\
    wghtd avg     &  0.9953   &      0.9906    &   0.9960 & 0.9209   \\
        \bottomrule
\end{tabular}
\end{table} 

The results demonstrate that the 1D-CNN significantly outperforms the LSTM model in the classification of arrhythmias across all evaluated performance metrics. The 1D-CNN achieves notably higher accuracy, with category-specific accuracies nearing 1.000, and superior precision and recall values, particularly excelling in the identification of false positives and ventricular non-sustained arrhythmias (FPNs and FVNs). The F1 score, a measure of the balance between precision and recall, further underscores the 1D-CNN’s robustness, with values consistently above 0.990. In contrast, the LSTM model, while still performing adequately, exhibits lower accuracy, precision, recall, and F1 scores, particularly in the detection of normal (N) and ventricular non-sustained arrhythmias (FVNs), suggesting that it may struggle with maintaining the balance between precision and recall. 

\begin{figure}[t]
\centering
\includegraphics[width=\columnwidth]{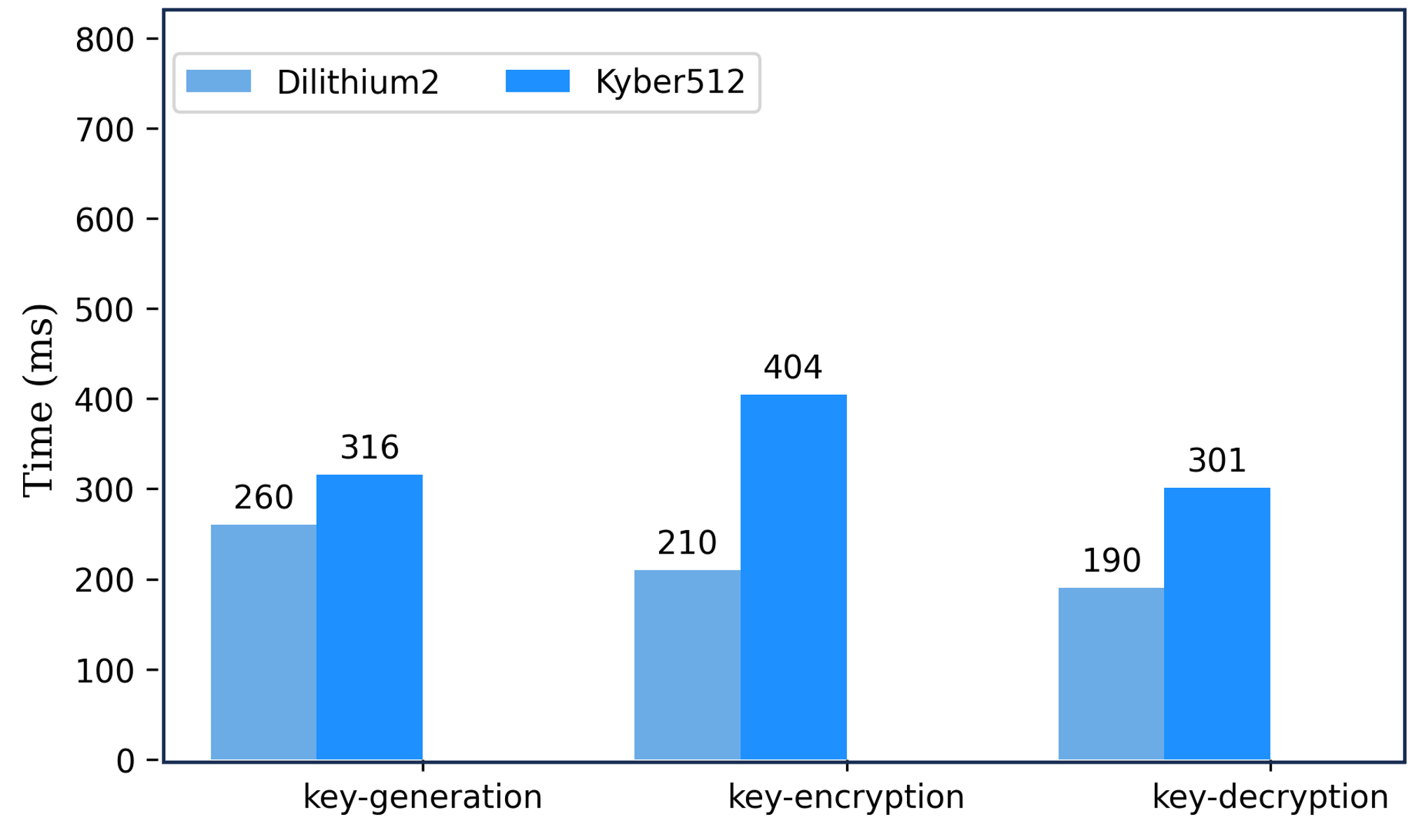}
\caption{Latencies in (ms) for kyber512 and dilithium2 proccess}
\label{fig:kyber}
\end{figure}

These findings indicate that the 1D-CNN is a more effective and reliable model for real-time arrhythmia classification, particularly in scenarios where the accurate and timely detection of irregular heart rhythms is critical. CNN operations can be applied across the entire sequence simultaneously, leading to faster processing times. This efficiency is crucial for real-time applications where quick decision-making is essential. While, sequential nature makes LSTMs computationally more intensive and slower compared to CNNs, particularly when dealing with long sequences, which can be a limitation in real-time scenarios. 

To assess the performance of the proposed network, we compared it to some state-of-the-art methods in the literature. We record the performance of the proposed network model (in bold) and some recent arrhythmia classiﬁcation using the MIT-BIH arrhythmia database in table ~\ref{tab:CNNrelatedwork} .From table ~\ref{tab:CNNrelatedwork}, it is evident that our proposed 1D-CNN achieved good performance while utilising less resources.

\begin{table*} [!h]
\centering
\caption{Comparison between the related work and the proposed 1D-CNN model}
\label{tab:CNNrelatedwork}
\small
\begin{tabular}[t]{@{}p{1.4cm}p{1.1cm}p{5.0cm}p{2.2cm}p{1.6cm}p{1.6cm}p{2.0cm}@{}}
\toprule
Ref. & Year & Classification Technique & Resource Cnsmptn & No. of Layers & No. of Classes & Accuracy
\\ \midrule
\midrule
\cite{lu2021kecnet} & 2021  &Neural network  & High  &  15 & 5 & 99.31\%
\\ 
\cite{cimen2021transfer} & 2021  &  Transfer Learning  & High &  18 &  2 & 90.42\%
\\
\cite{farag2022self} & 2022  & STFT-CNN  &  Moderate & ...... & 5 & 99.0\%
\\ 
\cite{banos2023novel} & 2023  & H-PSOCNN &  Moderate & ..... & 5 &  98.0\%
\\ 
\cite{xia2023transformer} & 2023 & CNN,DAE + Transformer  & Moderate   & 5 & 5 &  97.66\%
\\ 
\cite{islam2024cat} & 2024  & CNN,Attention + Transformer  &  High & 8 & 5 &  99.58\%
\\ 
\cite{alamatsaz2024lightweight} &  2024 & 1D-CNN+LSTM  & Moderate  & 11  &  9 &  98.24\%
\\ 
\cite{borra2024cadnet} & 2024  &  CAD-Net(1D‑CNN) &  Moderate & ... & 5 & 99.54\%
\\ 
\cite{venkatesh2024automated} & 2024  & 1DCNN-BiLSTM  &  Low & 7 & 5 & 93.7\%
\\ 
\textbf{This Study} & \textbf{2024}  & \textbf{Proposed 1D-CNN}  & \textbf{Low} & \textbf{2} & \textbf{5} &  \textbf{97.4\%}
 \\ \bottomrule
\end{tabular}
\end{table*}

\subsection{1D-CNN Comparison against different class configurations}

The performance metrics of the resource-optimized 1D-CNN model for arrhythmia classification across unbalanced, oversampled, and undersampled datasets reveal distinct trade-offs associated with each data sampling strategy. Table ~\ref{tab:classconfigurations} depicts the performance metrics of 1D-CNN for different diagnostic classes across various model configurations.

Using an unbalanced dataset, the model demonstrates strong overall performance, with high accuracy (ranging from 0.975 to 0.998), precision, and recall across all diagnostic classes. The F1 scores are also consistently high, indicating a well-balanced performance. For example, the recall for FPNs (0.995 ± 0.0140) and FVNs (0.997 ± 0.0080) is particularly impressive, suggesting that the model is adept at correctly identifying positive instances even with an unbalanced dataset.

\begin{table*}
\centering
\caption{1D-CNN Performance metrics for different diagnostic classes across various model configurations}
\small
\begin{tabular}{@{}p{0.10\linewidth}p{0.13\linewidth}p{0.20\linewidth}p{0.20\linewidth}p{0.20\linewidth}@{}}
\toprule
\textbf{Class} & \textbf{Metric} & \textbf{Unbalanced} & \textbf{Oversampled} & \textbf{Undersampled} \\ 

\midrule
\midrule

\multirow{5}{*}{\textbf{N}} & Accuracy & 0.975 $\pm$ 0.0030 & 0.973 $\pm$ 0.0030 & 0.951 $\pm$ 0.0035 \\ 
 & Precision & 0.983 $\pm$ 0.0030 & 0.995 $\pm$ 0.0030 & 0.975 $\pm$ 0.0030 \\ 
 & Recall & 0.9915 $\pm$ 0.0030 & 0.960 $\pm$ 0.0030 & 0.983 $\pm$ 0.0030 \\  
 & $F_1$ score & 0.982 $\pm$ 0.0030 & 0.995 $\pm$ 0.0020 & 0.973 $\pm$ 0.0020 \\ \hline

\multirow{5}{*}{\textbf{FPNs}} & Accuracy & 0.992 $\pm$  0.0030 & 0.990 $\pm$  0.0025 & 0.984 $\pm$ 0.0025 \\ 
 & Precision & 0.992 $\pm$  0.0105 & 0.990 $\pm$  0.0085 & 0.962 $\pm$  0.0085 \\ 
 & Recall & 0.995 $\pm$  0.0140 & 0.9455 $\pm$  0.0080 & 0.975 $\pm$  0.0080 \\ 
 & $F_1$ & 0.980 $\pm$  0.0085 & 0.990 $\pm$  0.0060 & 0.980 $\pm$  0.0060 \\ 
 \hline

\multirow{5}{*}{\textbf{PVCs}} & Accuracy & 0.998 $\pm$  0.0025 & 0.979 $\pm$  0.0025 & 0.962 $\pm$  0.0025 \\
 & Precision & 0.990 $\pm$  0.0115 & 0.990 $\pm$  0.0075 & 0.980 $\pm$  0.0075 \\  
 & Recall & 0.97 $\pm$  0.0095 & 0.995 $\pm$  0.0090 & 0.985 $\pm$  0.0090 \\ 
 & $F_1$ & 0.9840 $\pm$  0.0075 & 0.998 $\pm$  0.0060 & 0.973 $\pm$  0.0060 \\ 
 \hline

\multirow{5}{*}{\textbf{AP}} & Accuracy & 0.998 $\pm$  0.0015 & 0.991 $\pm$  0.0025 & 0.962 $\pm$  0.0015 \\ 
 & Precision & 0.988 $\pm$  0.0040 & 0.985 $\pm$  0.0025 & 0.978 $\pm$  0.0035 \\ 
 & Recall & 0.995 $\pm$  0.0015 & 0.991 $\pm$  0.0030 & 0.985 $\pm$  0.0030 \\
 & $F_1$ & 0.990 $\pm$  0.0020 & 0.980 $\pm$  0.0020 & 0.980 $\pm$  0.0020 \\ 
 \hline

\multirow{5}{*}{\textbf{FVNs}} & Accuracy & 0.997 $\pm$  0.0020 & 0.997 $\pm$ 0.0020 & 0.968 $\pm$  0.0020 \\ 
 & Precision & 0.989 $\pm$  0.0050 & 0.991 $\pm$  0.0055 & 0.974 $\pm$  0.0055 \\ 
 & Recall & 0.997 $\pm$  0.0080 & 0.992 $\pm$ 0.0075 & 0.984 $\pm$  0.0075 \\  
 & $F_1$ & 0.997 $\pm$  0.0045 & 0.998 $\pm$  0.0050 & 0.988 $\pm$  0.0050 \\ \bottomrule
\end{tabular}

\label{tab:classconfigurations}
\end{table*}
In an oversampled dataset, the model generally maintains high performance, but there are some notable changes. Precision tends to increase slightly, especially in the N class (from 0.983 to 0.995), suggesting that oversampling helps the model to reduce false positives. However, there is a drop in recall for some classes, such as FPNs (from 0.995 ± 0.0140 to 0.9455 ± 0.0080), indicating that while the model becomes more precise, it may miss more positive instances when trained on oversampled data. The F1 scores reflect this trade-off, with minor increases in classes where precision improves and decreases in those where recall drops. 

Using an undersampled dataset, the model shows a mixed performance when trained on undersampled data. While accuracy remains consistent, precision generally decreases, as seen in the PVCs class (from 0.990 ± 0.0115 to 0.939 ± 0.0075), indicating a higher rate of false positives. Conversely, recall remains relatively stable or even improves in some cases, such as in the N class (from 0.9915 ± 0.0030 to 0.993 ± 0.0030). This results in F1 scores that are generally stable or slightly lower compared to the unbalanced scenario, indicating that undersampling may help the model focus more on detecting positives but at the cost of precision.

The results demonstrate that while the 1D-CNN model performs robustly across all sampling strategies, trade-offs are depending on the data configuration. The unbalanced dataset provides the best overall performance with a good balance between precision and recall. Oversampling improves precision at the expense of recall, while undersampling can enhance recall but may reduce precision. The choice of sampling strategy should therefore align with the specific goals of the arrhythmia classification task, whether prioritizing the reduction of false positives or the capture of true positives.

\subsection{Resource, Security and Cost Analysis}
The proposed model demonstrates significantly lower resource consumption and implementation costs compared to recent studies. As shown in Table ~\ref{tab:CNNrelatedwork}, the model effectively classifies multiclass arrhythmia using only two convolutional layers, optimized for deployment on single-board computers. Its implementation with biosensors, integrated with raspberry pi and STM-based boards, offers a cost-effective solution for home monitoring applications. This makes it a practical complement to large-scale ubiquitous healthcare systems, contributing to medical history documentation and enabling real-time responses in time-sensitive scenarios.
\begin{table}[t]
    \centering
    \caption{Average Loading time of the different Arrhythmia classification requirements}
    \label{tab:loadinglibraries}
    \small
    \begin{tabular} {@{}p{0.58\linewidth}p{0.28\linewidth}@{}}
        \toprule
\textbf{Process} & \textbf{Execution Time} \\
        \midrule
        Loading Numerical Libraries &  960\~ms \\
        Loading Tensorflow and Keras & 1478\~ms \\
        Loading Trained Model & 6683\~ms \\
        \bottomrule
    \end{tabular}
\end{table}

In table ~\ref{tab:loadinglibraries} we illustrates the average loading time corresponding to the numerical libraries, deep learning libraries (Tensorflow and Keras), and the model at edge gateways. One of the important consideration in resource-constrained devices is latency during key generation, encryption, and decryption process of KEMs and digital signatures. Figure ~\ref{fig:kyber}, shows the latencies (ms) we got using \textit{kyber512} and \textit{dilithium2} using TLS-based secure communication over wi-fi. It shows the latencies in miliseconds we got using \textit{Kyber512} key generation \textit{(kk\_gen)}, \textit{Kyber512} key encryption \textit{(k\_enc)}, and keyber512 key decryption \textit{(k\_dec)}. Latencies using \textit{dilithium2} key generation \textit{(dk\_gen)}, \textit{dilithium2} key encryption \textit{(d\_enc)}, and \textit{dilithium2} key decryption \textit{(d\_dec)}. Results ensure that resource-constrained devices can leverage post-quantum cryptosystems effectively without compromising security or consuming excessive computational resources and time.

\vspace{12pt}
\section{Conclusion}

Our proposed \textit{EGBUH} framework extends the blockchain paradigm at EGs, eliminating the need for multiple layers of network infrastructure. By enabling resource-constrained computing devices to make autonomous decisions and process data at the edge, this approach eliminates the need for intermediaries. 

Furthermore, in addition to the continuous monitoring system for individuals, incorporating edge-level intelligence for rapid medical intervention. The system is designed to facilitate patient self-monitoring and preventive healthcare through the use of IoMT devices, employing computationally efficient methods. Processed medical data is securely stored using a private ethereum network, allowing for the sharing of anonymized information with legal guardians, healthcare providers, and researchers, while safeguarding patient privacy. The edge layer based on a single board computer is capable of processing privacy-critical sensitive information at the edge node. It ensures the user's privacy by discarding the raw data and only saving the processed information.

Firstly, SVM is employed for simple binary classification of abnormal signals. Additionally, a resource-efficient 1D-CNN is proposed for the multi-class classification of arrhythmias. Key contributions include the deployment of ML algorithms at EGs level for rapid anomaly detection and real-time detection and classification of arrhythmias using a two-channel ECG system. An extensive set of experiments and their detailed comparative analysis shows the viability of our proposed resource-optimized 1D-CNN in time-critical scenarios.
 
By filtering sensor data before it is written into the blockchain,it reduces the size and run it on gateway devices. With turing complete ethereum smart contracts, system decision-making, thresholds, and agreements are incorporated. It ensures prompt treatment by alerting specified devices via alert notification, hospital, or any predefined entity.

\bibliographystyle{IEEEtran}
\bibliography{main}
\end{document}